\documentclass[pra,twocolumn,superscriptaddress,10pt,floatfix]{revtex4-2}
\usepackage{amsmath}
\usepackage{hyperref}
\usepackage{graphicx}
\usepackage{xcolor}
\usepackage{epstopdf}
\usepackage{verbatim}
\usepackage{physics}
\usepackage{pgfplots}
\usepackage[normalem]{ulem}
\usepackage{siunitx}
\usepackage{orcidlink}
\usepackage{array}
\usepackage{tabularx}
\usepackage{booktabs}
\usepackage{float} 
\usepackage{xcolor}

\begin{document}

\title{Absolute frequency  measurement of the $^{40}$Ca$^{+}$ clock transition using a GNSS link to the SI second}
\author{M.~Guevara-Bertsch}
\affiliation{Institut f\"ur Quantenoptik und Quanteninformation, \"Osterreichische Akademie der Wissenschaften, Technikerstr. 21a, 6020 Innsbruck, Austria}
\affiliation{Universit\"at Innsbruck, Institut f\"ur Experimentalphysik, Technikerstr. 25, 6020 Innsbruck, Austria}
\affiliation{Alpine Quantum Technologies GmbH, 6020 Innsbruck, Austria}

\author{ M.~K~Joshi}
\altaffiliation{Present address: Singapore University of Technology and Design, 8 Somapah Road, 487372, Singapore}
\affiliation{Institut f\"ur Quantenoptik und Quanteninformation, \"Osterreichische Akademie der Wissenschaften, Technikerstr. 21a, 6020 Innsbruck, Austria}

\author{ M.~I~Hussain}
\altaffiliation{Present address: Qatar Center for Quantum Computing, College of Science and Engineering, Hamad Bin Khalifa University, Doha, Qatar}
\affiliation{Institut f\"ur Quantenoptik und Quanteninformation, \"Osterreichische Akademie der Wissenschaften, Technikerstr. 21a, 6020 Innsbruck, Austria}

\author{R.~Blatt}
\affiliation{Institut f\"ur Quantenoptik und Quanteninformation, \"Osterreichische Akademie der Wissenschaften, Technikerstr. 21a, 6020 Innsbruck, Austria}
\affiliation{Universit\"at Innsbruck, Institut f\"ur Experimentalphysik, Technikerstr. 25, 6020 Innsbruck, Austria}

\author{C.~F.~Roos}
\affiliation{Institut f\"ur Quantenoptik und Quanteninformation, \"Osterreichische Akademie der Wissenschaften, Technikerstr. 21a, 6020 Innsbruck, Austria}
\affiliation{Universit\"at Innsbruck, Institut f\"ur Experimentalphysik, Technikerstr. 25, 6020 Innsbruck, Austria}

\begin{abstract}
We report the absolute frequency measurement of the $4s$ $ ^{2}S_{1/2}\leftrightarrow 3d$ $^{2}D_{5/2}$ $^{40}$Ca$^{+}$ clock transition with respect to the SI second.  To perform this measurement, a link between our laboratory in Innsbruck and the clocks realizing the Coordinated Universal Time at the Physikalisch-Technische Bundesanstalt (PTB) in Braunschweig was installed and characterized using the Global Navigation Satellite System GNSS. The comparison between our clock and the ones at PTB was done using the Precise Point Positioning  technique. After the evaluation of the systematic shifts, the measured transition frequency is 411 042 129 776 400.4$\pm$0.6 Hz with a fractional uncertainty of 1.5 $\times$ 10$^{-15}$. 
The stability of the clock measurements was also corroborated by comparing two different calcium ion clock experiments, which share the clock laser source at our institute. Furthermore, after careful evaluation of the trap-drive induced ac magnetic fields, we estimate ac Zeeman shifts on the $D_{5/2}$ sublevels and reevaluate the Land\'e g-factor of the $3d$ $^{2}D_{5/2}$ level to be g$_{5/2}= 1.200329(1)$. 
\end{abstract}


\maketitle

Optical atomic clocks based on singly-charged ions, with one or two valence electrons, such as Ca$^{+}$, Sr$^{+}$, Hg$^{+}$, Al$^+$, Lu$^+$ have been extensively studied due to their relatively simple electronic level structure \cite{udem2001absolute,chwalla2009absolute,margolis2004hertz, zhang2025absolute, brewer2019al}. These species have the combined advantages of exhibiting an excellent performance of frequency uncertainty and instability, with transitions that can be accessed with commercially available laser sources and high-quality optics. Through the implementation of simple schemes probing multiple Zeeman transitions, the effects of systematic shifts such as the quadrupole, linear Zeeman and ac Zeeman shifts can be completely canceled from the determination of the clock transition. Additionally, for trapped-ion clocks having transitions exhibiting a negative differential scalar polarizability such as Ca$^{+}$ or Sr$^{+}$, the effects due to the thermal secular motion of the ion and the uncompensated static electric fields that push the ion away from the rf null position can be strongly reduced by an appropriate choice of the rf drive frequency \cite{berkeland1998minimization, dube2013evaluation}.

Ca$^{+}$ ion clocks have reached fractional uncertainties on the order of 
$10^{-17}$ \cite{zhang2023absolute, sokolik2026direct} and a clock with systematic uncertainties below $10^{-18}$ has been demonstrated \cite{Zhang2026}. Since its first measurement in 2009 \cite{chwalla2009absolute}, the absolute frequency  of the  $4s$ $ ^{2}S_{1/2}\leftrightarrow 3d$ $^{2}D_{5/2}$  $^{40}Ca^{+}$ clock transition has been reported four more times: at the National Institute of Information and Communications (NICT)  in Japan \cite{matsubara2012direct} and at the Wuhan Institute of Physics and Mathematics (WIPM) \cite{huang2012hertz,huang2017comparison,huang2016frequency}. The earlier published results (between 2009 and 2012) are in disagreement with the latest measurements (between 2013 and 2025). We present the results of a new campaign to measure the absolute frequency of the  $4s$ $ ^{2}S_{1/2}\leftrightarrow 3d$ $^{2}D_{5/2}$  $^{40}Ca^{+}$ clock transition with respect to the coordinated universal time UTC(PTB) at the Physikalisch-Technische Bundesanstalt by means of a GNSS link using the Precise Point Positioning (PPP) technique. The campaign was held for 10 days from the 16$^{th}$ to the 25$^{th}$ of June 2021. After the evaluation of the systematic shift the transition frequency is measured as 411 042 129 776 400.4$\pm$0.6 Hz with a fractional uncertainty of 1.5 $\times$ 10$^{-15}$. 

Following an approach similar to the one described in Ref.\cite{chwalla2009absolute} we implement a probing scheme with six transitions in order to cancel the effects of the linear Zeeman, quadrupole and ac-Zeeman shift in the determination of the clock transition frequency. However, the model implemented to estimate the error budget during this campaign differs from the one used in Ref.\cite{chwalla2009absolute} by taking into consideration the effects of oscillating magnetic-fields arising from ion trap-induced rf current in electrodes. Even if the effects of ac Zeeman shifts cancel out in the determination of the clock transition their effect needs to be considered in the determination of the Land\'e g-factor of the D$_{5/2}$ level~\cite{gan2018oscillating}. Following the characterization of the oscillating-magnetic-field arising from trap-induced rf currents in the electrodes, described in \cite{joshi2024characterization}, we have incorporated the ac-Zeeman shift to each probed transition to perform a new revised measurement of the Land\'e g factor and measure a value of g$_{5/2}= 1.200329(1)$. This value, mainly limited by the uncertainty in the estimation of the oscillating-magnetic field, differs from the one reported in Ref.\cite{chwalla2009absolute} at the $10^{-6}$ level. This is probably due to the fact that during the 2009 campaign the effects of the ac Zeeman shifts were not considered in the determination of the Land\'e g-factor of the D$_{5/2}$ level. Our measurements are in agreement with the value reported in Ref.~\cite{Ma:2024, McMahon2026} using measurements that also take into consideration the effect of trap-induced ac-Zeeman shifts.

The paper is organized as follows: Section \ref{Sec:setup} focuses on the description of the experimental setup. Section \ref{sec:FreqMeasurement}  presents the frequency measurement of the clock transition via the GNSS link. The evaluation of the systematic shifts and measurement uncertainties are described in section \ref{Sec:errorbudget}. Additionally, the effects of oscillatory magnetic fields are introduced for a new determination of the Land\'e g factor of the D$_{5/2}$ level. Finally, the last section summarizes the results of the absolute frequency measurement.
\section{Experimental Setup \label{Sec:setup}}

The ion trap and laser system have been previously described in detail in Refs.~\cite{Guggemos:2015,guggemos2019frequency}. A single $^{40}$Ca$^{+}$ is confined in a blade-style linear Paul trap with oscillation frequencies of about 1.5~MHz in the radial directions and 820~kHz in the axial direction. The trap is mounted inside an ultra-high vacuum chamber at a residual gas pressure of roughly 10$^{-11}$\,mbar. To define the quantization axis a magnetic field of 3.0787\,G is created by an assembly of permanent magnets mounted in two concentric permanent magnet holders on both sides of the vacuum system, with quantization axis parallel to the trap axis. 
\begin{figure}[t]
\includegraphics[width=1\linewidth]{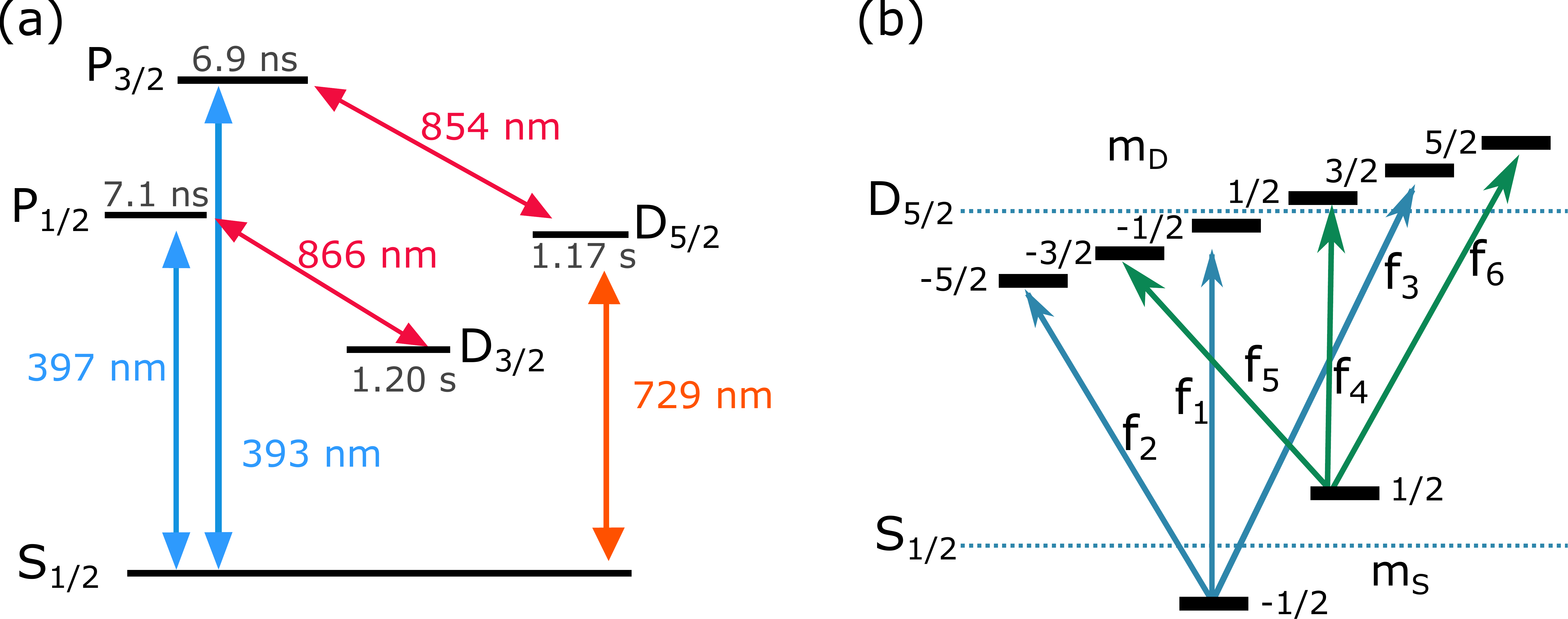}
 \caption{\label{fig:energylevels}(a)~Reduced energy level scheme of $^{40}$Ca$^+$. The S$_{1/2}\leftrightarrow$D$_{5/2}$ quadrupole transition at 729~nm is used as a clock transition. Doppler cooling and state detection are carried out on the cycling transition S$_{1/2}\leftrightarrow \textrm{P}_{1/2}$ at 397~nm. Two infrared lasers at 866~nm and 854~nm pump out the metastable D-states to prevent population trapping. (b)~Transitions f$_{1}$, f$_{2}$, f$_{3}$, f$_{4}$, f$_{5}$, f$_{6}$ are probed in order to cancel the linear Zeeman shift, ac-Zeeman shift and the quadrupole shift.
 } 
\end{figure}
The ions are loaded into the trap by two-step photoionization from neutral atoms generated via short-pulse ablation from a target located 26~mm above the trap axis. Figure~\ref{fig:energylevels}~(a) shows the energy levels of the $^{40}$Ca$^{+}$ that are relevant for the operation of the ion clock. For Doppler cooling, state detection and optical pumping we use the light of a frequency-doubled extended-cavity diode laser at 397~nm resonant with the $\mathrm{S}_{1/2}\leftrightarrow \mathrm{P}_{1/2}$ transition. The D$_{3/2}$ and  D$_{5/2}$ states are pumped out by means of extended-cavity diode lasers resonant with the D$_{3/2}\leftrightarrow$ P$_{1/2}$ (at 866~nm) and the D$_{5/2}\leftrightarrow$ P$_{3/2}$ (at 854~nm). The fluorescence emitted by calcium ions at 397~nm is recorded by a photomultiplier tube and an electron-multiplying CCD camera.

The $\mathrm{S}_{1/2}\leftrightarrow \mathrm{D}_{5/2}$ quadrupole transition at 729\,nm is probed by a titanium-sapphire laser locked to a high-finesse cavity yielding a spectral linewidth below 10\,Hz on the time scale of a few seconds. The probe light is measured using a femtosecond frequency comb referenced to a passive hydrogen maser. Following a procedure similar to the one described in Refs.~\cite{ray2005geodetic,dube2017absolute}, the maser is calibrated with respect to UTC(PTB) via a GNSS satellite intercomparison using the Natural Resources Canada Precise Point Positioning (CSRS-PPP) data-processing technique \cite{Orgiazzi2005}. Combining the results of the intercomparison with the information from the Bureau International des Poids et Measures (BIPM) circular-T reports the maser is finally calibrated in its absolute offset from the SI second.  Our GNSS link is composed of a Septentrio PolaRx4TR receiver with NovAtel’s GNSS-750 antenna and a passive hydrogen maser VCH-1008 from the company T4 Science as external time and frequency reference. 

To measure the clock transition, six Zeeman transitions shown in Fig.~\ref{fig:energylevels}~(b) are probed by Ramsey experiments with probe times of 5~ms. The center frequency of each transition is determined with two measurements, where the phase of the second Ramsey pulse is shifted by $\pm \pi/2$ with respect to the phase of the first one, respectively.
From these measurements, the magnetic field strength and the laser frequency relative to the ion are inferred. This information is fed back to the frequency of an acousto-optical modulator (AOM) to compensate for slow drifts of the reference cavity and magnetic field.

Each probing cycle consists of 5~ms of Doppler cooling on the $S_{1/2}\leftrightarrow P_{1/2}$ transition at a wavelength of 397~nm, repumping at 866~nm to counter decay into the $D_{3/2}$ level and repumping at 854~nm to clear out population
from the D$_{5/2}$ level. State initialization is done by optical pumping into the $|S_{1/2},m_{j}=\pm1/2\rangle$ level, using two different methods consecutively: one using circularly polarized light at 397 nm to drive a $\sigma\pm$ transition between S$_{1/2}$ and P$_{1/2}$ followed by the second one using the narrowband laser at 729 nm to selectively couple the $|S_{1/2},m_{j}=\pm1/2\rangle$ level to the D$_{5/2}$ manifold, from where it is pumped out using light at 854 nm. During the pumping cycle, repumper light at 866 nm ensures that decays to the D$_{3/2}$ state are returned to the pumping cycle. Optical pumping is followed by the probing pulses at 729~nm and finalized by the 5~ms ion-state detection period. The polarization and angle of the 729~nm laser beam is adjusted to obtain a similar coupling strength for all 6 transitions \cite{roos2000controlling}. The experiments were performed with a Ramsey probe time $\tau_{R}$ of 5~ms and  a $\pi/2$ pulse length of approximately 100~$\mu$s for each probed transition. Every cycle is repeated 100 times for each individual frequency measurement. In total, it took 30~s for probing all six transitions in one measurement cycle.

\section{Frequency measurement via GNSS link \label{sec:FreqMeasurement}}
The laser frequency is measured via an optical beat note with the light of the frequency comb. All ``out-of-lock'' events where any of the recorded frequencies deviate by more than three standard deviations from the resulting mean value are discarded. To detect cycle slips, the signal from the optical beat note is split into two parts, one of which is attenuated by 3~dB. They are both detected by two separate counters. Following the approach described in \cite{droste2015characterization}, all data points for which the counted signals disagree by more than 8 times the median absolute deviation (MAD) are discarded. The measurements of the laser frequency with the comb are combined with the deviations of the laser frequency relative to the ion's one extracted from the Ramsey phase experiments over the 6 transitions. The valid points measured with the frequency comb during the time window in which the ion data is taken are averaged. The data extracted directly with the ion is then combined with the frequency comb data.

The GNSS receiver is referenced to the 10~MHz and the pulse per second (PPS) signal from the passive hydrogen maser. The receiver decodes the signal from the GNSS satellites and generates the receiver independent exchange format (RINEX) files. Using the NRCan Precise Positioning software, the time delay between the maser and the satellites' clocks is estimated. A similar setup located at PTB estimates the time delay between the satellites' clocks and the UTC(PTB). Combining the two data sets, the time delay of the hydrogen maser with respect to the UTC(PTB) is estimated. From these measurements, we calculated the relative frequency offset of the passive hydrogen maser with respect to the UTC(PTB). The data is recorded every 30~s and processed in 5~min batches. 

The frequency measurements relative to the hydrogen maser are combined with the characterization of the hydrogen maser with respect to the UTC(PTB), to finally obtain the frequency of the calcium clock transition with respect to the UTC(PTB). The valid points measured with the frequency comb and the ion trap during the time window used for the PPP analysis are averaged and combined with the calibrated maser data. 
\begin{figure}
\includegraphics[width=\columnwidth]{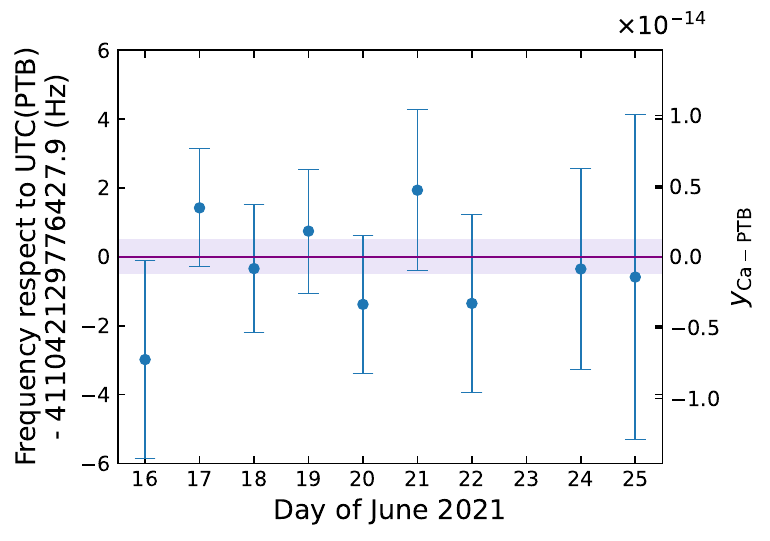}
 \caption{\label{fig:frequencymeasurement}{\textbf{Absolute frequency measurements referenced to UTC(PTB).} Data points with error bars represent mean absolute frequency of specified days. The mean absolute frequency estimated from all measurements is 411 042 129 776 427.9~Hz (solid line) with an uncertainty of 0.6~Hz (shaded area). The fractional frequency deviation is given on the right side.}
}
\end{figure}

Figure~\ref{fig:frequencymeasurement} shows the frequency measured on each day of the measurement campaign. The total mean is 411 042 129 776 427.9~Hz, with a statistical uncertainty of 0.6~Hz, represented by the purple shaded region in the plot. The error bars in the plot represent the daily statistical uncertainty, limited mainly by out-of-lock events. 
\begin{figure}
\includegraphics[width=\columnwidth]{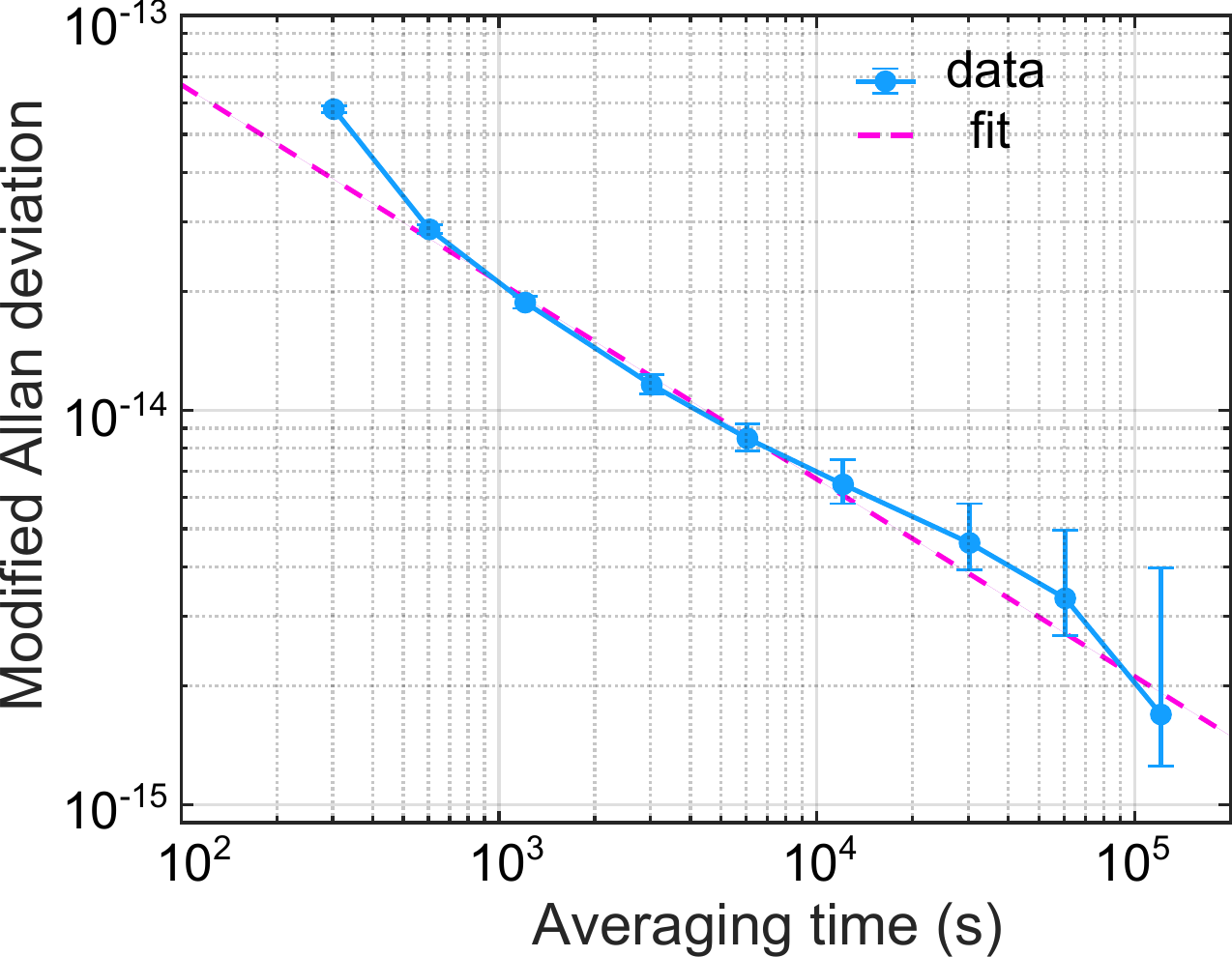}
 \caption{\label{fig:allanUTC}\textbf{Fractional 
 modified Allan deviation plot of the frequency comparison.} The data points are presented in blue. The dotted pink line is a fit that follows $\sigma_y(\tau)=6.83\times 10^{-13} \tau^{-1/2}$.}
\end{figure}
Figure~\ref{fig:allanUTC}  shows the fractional modified Allan deviation plot of the total measurement campaign. The instability reaches a minimum at low $10^{-15}$ values at an averaging time of $10^{5}$ s. A logarithmic fit of the  modified Allan deviation data with respect to the averaging time shows a stability close to $6.83 \times 10^{-13} \tau^{-1/2}$ revealing the presence of white frequency noise most likely originating from the hydrogen maser. From these measurements, the frequency of the clock transition, referenced to the UTC(PTB), without corrections for systematic shifts, is determined to be 411 042 129 776 427.9$\pm$ 0.6~Hz.  

\begin{figure}[h]
\begin{center}
\includegraphics[width=\columnwidth]{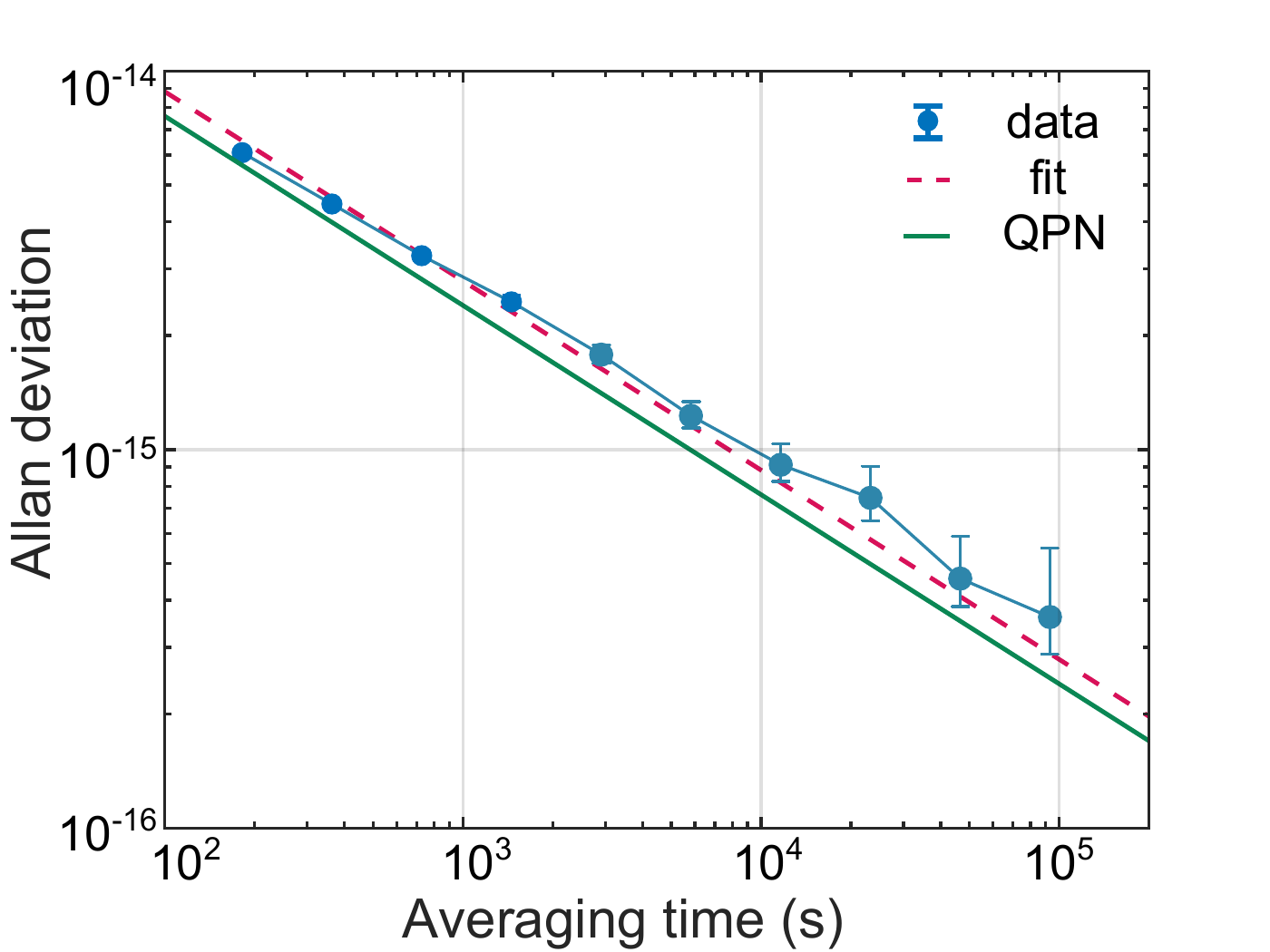}
 \caption{\label{fig:allan2labs}\textbf{Standard fractional Allan deviation plot of the frequency comparison of two $^{40}$Ca$^{+}$ clock divided by $\sqrt{2}$.} The data points are presented in blue. The dotted red line is a fit that follows $\sigma_y(\tau)=8.7\times 10^{-14} \tau^{-1/2}$. The green solid line is theory calculation that follows $\sigma_y(\tau)=7.4\times 10^{-14} \tau^{-1/2}$.}
\end{center}
\end{figure}

To determine the stability of our spectroscopic measurements independently from the hydrogen maser, we made a frequency comparison between our $^{40}$Ca$^{+}$ frequency reference and a similar one located in a neighboring laboratory that hosts a similar setup dedicated to experiments with calcium ion chains, described in detail in Ref.~\cite{hempel2014digital}. In both setups, ions were probed with the same laser light that was split into two beams. Each laser beam was sent to the respective experimental setup via optical-pathlength-stabilized fibers. Both experiments were carried out with a probe time $\tau_{R}$ of 5000~$\mu$s and $\pi/2$ pulse lengths of approximately 100~$\mu$s for each transition. 
For a continuous data set taken over 30 hours, the Allan deviation of the frequency difference between the two traps was evaluated. The results are plotted in Figure~\ref{fig:allan2labs}. The individual contribution of each clock to the stability can be considered equivalent, so in order to get the stability for a single clock, the Allan deviation is divided by $\sqrt{2}$. A fit to the measured data reveals that the single clock stability is close to $8.7\times 10^{-14}~\tau^{-1/2}$.
The theoretical Allan deviation (assuming only quantum projection noise) as a function of the averaging time $\tau$ obtainable with Ramsey experiments using our experimental parameters \cite{peik2005laser} is  equal to approximately $7.4\times 10^{-14} \tau^{-1/2}$ and illustrated as a solid green line in Fig.~\ref{fig:allan2labs}. The small deviation of the fitted trend from the theory prediction might be caused by the parameters of the servo algorithm that stabilizes the laser frequency to the single-ion signal and that eliminates errors due to laser frequency drift or by some external yet to be defined noise sources. However, these results allow us to confirm that our spectroscopy measurements are mostly limited by quantum projection noise. 

\section{Evaluation of systematic frequency shifts and measurement uncertainties\label{Sec:errorbudget} }
The implementation of the two-phase symmetric Ramsey probing scheme on six transitions, previously described and illustrated in Fig.~\ref{fig:energylevels}(b) ensures that the linear Zeeman, quadrupole and ac-Zeeman shifts are canceled even in the presence of magnetic or electric field drifts. However, the scheme does not completely cancel the shift in the presence of non-linear temporal variations of the magnetic field. Nonetheless, it considerably minimizes its effects. Using the spectroscopy data from all probed transitions, the variations of the magnetic field and electric field gradient are constantly monitored. We observe a slow linear drift of the magnetic field $(-0.06\pm 2)$~nG$/$s that most likely originates from temperature variations of the permanent magnets. The maximum quadratic variation of the magnetic field measured is on the order of 0.002~nG/s$^{2}$ which corresponds to a linear Zeeman shift of 1~$\mu$Hz. A very slow linear drift of the electric field gradient equal to ($0.1\pm0.3)$~$\mu$Hz$/$s is measured with no visible higher order variations. 

In the presence of the magnetic field, the magnetic dipole interaction perturbatively couples the D$_{3/2}$ and D$_{5/2}$ Zeeman sublevels with a magnetic quantum number $\lvert m_{j} \rvert \leq 3/2$ to each other, inducing a second-order Zeeman shift \cite{dube2013evaluation}. The shifts of the $D_{5/2}$ levels are not symmetric with respect to the baseline at zero magnetic field and are therefore not canceled by our probing scheme. The contribution of this shift can be estimated with very high precision, given that its uncertainty is mainly limited by the determination of the magnetic field.  The second order Zeeman shift during the measurement campaign is calculated to be ($1.366850\pm 5\times 10^{-6}$)~Hz. 

Our model also includes effects of oscillating magnetic fields that arise from the imbalanced rf currents in the trap electrodes. An ac-magnetic field with a component (B$_{\perp}$) perpendicular to the dc-magnetic field setting the quantization axis can drive spin flip transitions between neighbouring Zeeman states \cite{gan2018oscillating}. If the field is off-resonant from the transition, ac-Zeeman shifts occur. These shifts are dependent on the magnitude of B$_{\perp}$. For a given level $\ket{J, m}$ the ac-Zeeman shift can be calculated\cite{joshi2024characterization}, yielding
\begin{equation}
\delta_{ac} = -\frac{(g \mu_{\textrm{B}}\textrm{B}_{\perp})^{2}}{8\hbar^{2}} \left( \frac{1}{\Omega_{\textrm{RF}}-\Delta}- \frac{1}{\Omega_{\textrm{RF}}+\Delta}\right)m,
\label{eq:aczeeman}
\end{equation}
where g is the Land\'e g-factor and $\Delta$ is the Zeeman splitting of a given level, denoting g$_{S}$ and $\Delta_{S}$  for the S$_{1/2}$ and g$_{D}$ and $\Delta_{D}$ for D$_{5/2}$, respectively for the involved atomic levels. $\Omega_{\textrm{RF}}$ is the frequency of the rf source.
The formula shows that the ac-Zeeman shift is proportional to the square of the rf-magnetic field, but linear in the magnetic quantum number. Thus, our probing scheme ensures that these shifts are cancelled in the determination of the clock transition frequency. However, these shifts need to be taken into consideration for the determination of the Land\'e g-factor of the D$_{5/2}$ level. The transversal rf field B$_{\perp}$ in our setup was determined to be equal to 54$\pm$9~mG (67$\pm$9~mG at a higher rf power see Ref.~\cite{joshi2024characterization}). Including the linear Zeeman, second order Zeeman, quadrupole and ac zeeman shift with the individual transition frequency data combined with the Land\'e g-factor of the S$_{1/2}$ level \cite{tommaseo2003factor} we determined the g-factor of the D$_{5/2}$ level to be equal to g$_{D}$=1.200329(1). This value differs from the previously measured value of g$_{D}$, as reported in Ref. \cite{chwalla2009absolute} however is in agreement with recent results \cite{Ma:2024,McMahon2026}. 
This discrepancy is attributed to the AC Zeeman effect, giving rise to a fractional change of the Zeeman sublevel spacing on the D$_{5/2}$ level by $1.6\times 10^{-6}$ and on the S$_{1/2}$ level by $4.5\times 10^{-6}$. The uncertainty in the determination of the g-factor of the D$_{5/2}$ state is limited by the uncertainty in the determination of the transversal component of the ac magnetic field.

Another source of systematic shift is associated with the motion of a trapped ion. The ion motion, on one hand, gives rise to a second order Doppler shift, while, on the other hand its interaction with the 
electric field leads to a quadratic Stark shift \cite{berkeland1998minimization, dube2013evaluation}. Ion motion can be divided into three categories. The first is secular motion, which is related to the thermal energy of the ion. The second is intrinsic micromotion (IMM), which arises from a non-vanishing amplitude of secular motion. The third is excess micromotion (EMM), which is relevant when uncompensated dc fields push an ion away from the RF null line. For ions with a negative differential scalar polarizability such as $^{40}$Ca$^{+}$ and $^{88}$Sr$^{+}$ the combined effect of these two shifts cancels out at the magic RF drive frequency \cite{berkeland1998minimization,huang2019ca+} for excess micromotion and strongly reduces the combined shifts from intrinsic micromotion. For the $^{40}$Ca$^+$ ion, the exact magic RF drive frequency and the differential scalar polarizability were previously measured in Ref.~\cite{huang2019ca+} to be $\Omega_{\textrm{magic}}= 2\pi \times (24.801 \pm 0.002)$~MHz and $\Delta \alpha_{0} = (-7.2677 \pm 0.0021)\times 10^{-40}$~Jm$^{2}$V$^{-2}$. During our measurement campaign, the trap frequency was set to $\Omega = 2\pi \times 32$~MHz. Although our rf-drive frequency did not match the magic frequency, the combined frequency shift was nevertheless substantially reduced compared to the magnitude of the individual shifts.

We first discuss the shift that originates from the secular motion and IMM and contributes through second-order Doppler shifts. Since the second-order Doppler shift depends on the square of the respective velocity components, the shifts from both IMM and secular motion are proportional to the square of the secular motion's amplitude and thus depend linearly on the ion's temperature. To extract the temperature of the ion, we evaluated the phonon distribution $\overline{n}_{i}$ in each direction $i$ and obtained
a temperature of T$=(1.6 \pm 0.4)$~ mK. The combined second order Doppler shift due to secular motion and IMM  is equal to $\Delta \nu_{D2(sec+IMM)}= (-3.8\pm 0.7)$~mHz. Secondly, both, secular and IMM contribute to the quadratic stark shift by moving the ion into regions of non-zero electric rf-fields.
This contribution also depends upon the amplitude of motion hence can be ultimately expressed as a function of temperature. The combined quadratic Stark shift thus depends upon the trapping parameters and atomic properties, i.e. trap frequency, temperature of the ion and atomic polarizability etc. For the present case, this shift is dominated by secular motion and is estimated to be $(5.8\pm 1)$~mHz. 

Uncompensated static electric stray fields push the ion away from the RF null position, adding another source of second order Doppler and micromotion induced Stark shifts. The effects of these fields are minimized by shifting the ion position back to the RF null with two pairs of compensation electrodes that allow movement of ions in two orthogonal axes that are perpendicular to the principle axis of the trap. The magnitude of the excess micromotion, in the limit of low modulation, $\beta_{i}$ is determined from the intensity ratio of the micromotion sideband to the carrier of one of the Zeeman transitions \cite{berkeland1998minimization}. With the minimization of the micromotion the quadratic Stark shift is $\Delta\nu_{scalar,Excess}=(0.3\pm 0.1)$~mHz and the second order Doppler shift is $\Delta \nu_{D2,Excess}=(-0.18\pm0.08)$~mHz. The combined excess micromotion shift is equal to $\Delta \nu_{Excess}=(0.1\pm 0.1)$~mHz. The uncertainty of the combined shifts accounts for the contribution of each shift calculated separately.

The thermal electromagnetic radiation surrounding the ion shifts the energy of the two clock states by off-resonant coupling to other energy levels \cite{itano1982shift}. The stability and accuracy of the environmental temperature measurement limits the uncertainty of the black body radiation shift \cite{huang2012hertz}. The temperature generated at the trap as a function of the RF power was determined using the 4-point method. A detailed description of these measurements can be found in Ref~\cite{guggemos2017precision}. At the input power of 1~W the trap heats up approximately by (1.4$\pm0.5$)~K, while the environmental temperature of the laboratory, constantly monitored, remains constant at (294$\pm2$)~K. Assuming a uniform temperature distribution and a conservative 2~K uncertainty in the determination of the temperature, the black body radiation shift is equal to $\Delta\nu_{BBR}=(355\pm10)$~mHz. 

Another source of AC Stark shifts on the clock transition comes from the electromagnetic radiation of the lasers necessary for cooling, optical pumping and probing. In a Ramsey experiment, ideally all lasers are switched off during the waiting time and only the 729~nm laser is activated during the two probing pulses. However, since this cannot be done perfectly, the residual light fields can cause AC Stark shifts. Any shifts induced by the cooling and repumping lasers are most likely caused by unwanted laser light reaching the ion trap. Such leakage can occur either from the zeroth-order beam of the AOMs, if it is not sufficiently spatially separated at the input to the trap fiber, or from other undesired diffraction orders or scattered light originating from the AOMs if the radio frequencies are not properly shut. The optical path for each of these lasers is composed of a double pass and a single pass AOM assuring, when necessary, a combined attenuation of approximately 80~dB. The dominant shift for the 397~nm laser is generated by off-resonant coupling light to the cooling transition (S$_{1/2}\leftrightarrow$P$_{1/2}$). For both the 854~nm and 866~nm the dominant shift is generated by off resonant coupling to the D$_{5/2}\leftrightarrow$P$_{3/2}$ transition. To characterize any possible leaking light we carefully measured the intensity of each laser beam in front of the fiber that goes to trap in the ``off" condition. We measured powers that are equal to or lower than 0.1~nW which are limited by the resolution of our power meter. If we assume that only 1$\%$ of the unfocused and zeroth order light is coupled into the fiber and reaches the trap, we can estimate that any possible AC Stark shift generated by leaking light of the cooling and repumping lasers is going to be below  1~mHz. Although the exact origin and spectral content of the leaked light are unknown, we perform a worst-case estimate of the AC Stark shift by assuming that the leaked light is near resonance with the atomic transition, with the dominant contribution corresponding to twice the first-order frequency shift of the double-pass AOM. This conservative estimate yields an AC Stark shift below 20~mHz.

During the probing pulse, the 729~nm laser light contributes to the AC Stark shift by off-resonantly coupling to the Zeeman levels of the S$_{1/2}\leftrightarrow$D$_{5/2}$ as well as to the S$_{1/2}\leftrightarrow$P$_{1/2}$,  S$_{1/2}\leftrightarrow$P$_{3/2}$ and D$_{5/2}\leftrightarrow$P$_{3/2}$ transitions. This can happen either by the light resonant to the probed transition or by 0$^{\textrm{th}}$ order leaking light detuned by the corresponding single and double pass AOM frequencies. 
The shift generated by off resonant coupling with the Zeeman levels can be estimated as a function of the detuning $\Delta$ and the coupling strength $\Omega$ with respect to the probed transition. The 729~nm probing beam was adjusted to obtain a similar coupling strength for the $\Delta m =0,\pm1,\pm2$ transitions of approximately $\Omega=2\pi\times 1.250$~kHz. By probing 6 symmetric transitions with a similar coupling strength, the shift generated by the "on-resonance" probing pulses light cancels out. The shift corresponding to the 0$^{\textrm{th}}$ order leaking light is also minimized although not completely canceled, a conservative estimate is calculated considering the extra detuning of 620~MHz and a beam power of 0.1~nW. 
The shift generated by the off resonant coupling to the S$_{1/2}\leftrightarrow$P$_{1/2}$,  S$_{1/2}\leftrightarrow$P$_{3/2}$ and D$_{5/2}\leftrightarrow$P$_{3/2}$ for both probing time and waiting time are listed in table~\ref{tb:starkshift}.

\begin{table}
    \centering
    \setlength{\tabcolsep}{3pt}
    \renewcommand{\arraystretch}{1.1}
    \begin{tabular}{c c c c c}
        \hline
        \textbf{Radiation} &
        \begin{tabular}{c}
            \textbf{Shift}\\
            \textbf{wait time}\\
            \textbf{(mHz)}
        \end{tabular} &
        \begin{tabular}{c}
            \textbf{Shift}\\
            \textbf{probing}\\
            \textbf{(mHz)}
        \end{tabular} &
        \begin{tabular}{c}
            \textbf{Clock}\\
            \textbf{shift}\\
            \textbf{(mHz)}
        \end{tabular} &
        \begin{tabular}{c}
            \textbf{Error}\\
            \textbf{(mHz)}
        \end{tabular}
        \\
        \hline
        397~nm & $<1$ & $<0.03$ & $<1$ & $<1$ \\
        866~nm & $<1$ & $<0.03$ & $<1$ & $<1$ \\
        854~nm & $<1$ & $<0.03$ & $<1$ & $<1$ \\
        729~nm & $<0.1$ & $6$ & $6$ & $<0.1$ \\
        \hline
        \textbf{Total} & \textbf{3} & \textbf{6.1} & \textbf{9} & \textbf{2} \\
        \hline
    \end{tabular}
    \caption{
        \textbf{AC Stark shifts from cooling, repumping and probing lasers.}
        A conservative estimation of the shift and respective error for each
        of the lasers considering their effect during the probing pulses and
        the waiting time. Here, we assume that the corresponding laser fields
        are red detuned by twice the AOM drive frequency from the respective
        resonances.
    }
    \label{tb:starkshift}
\end{table}

Table~\ref{tb:errorbudget} summarizes the evaluation of the systematic shifts and uncertainties for the clock measurement. The total shift, considering only the spectroscopic measurements, is of 1.733~Hz with a fractional uncertainty of $2.5\times10^{-17}$. Our measurements are mainly limited by the determination of the blackbody radiation shift. To improve the precision of this measurement it is necessary to refine the determination of the ion's environmental temperature. Some groups have proposed the implementation a thermal imaging camera through a MgF$_{2}$ camera window to estimate the temperature at the Kelvin level as well as the implementation of temperature-stabilized boxes with chillers that could reduce the uncertainty to the 10$^{-18}$ level \cite{huang2020geopotential,huang2017comparison}. Other groups have implemented cryogenic chambers, greatly reducing both the shift and the uncertainty \cite{ushijima2015cryogenic,middelmann2012high}. The implementation of such techniques in the development of new optical ion trap experiments with calcium ions could help us push the uncertainty further down.

\begin{table*}
    \centering
    \small
    \setlength{\tabcolsep}{8pt}
    \renewcommand{\arraystretch}{1.05}

    \begin{tabular}{cccc}
        \toprule

        \textbf{Effect} &
        \begin{tabular}{c}
            \textbf{Clock}\\
            \textbf{shift (mHz)}
        \end{tabular}
        &
        \begin{tabular}{c}
            \textbf{Error}\\
            \textbf{(mHz)}
        \end{tabular}
        &
        \begin{tabular}{c}
            \textbf{Fractional}\\
            \textbf{error}
        \end{tabular}
        \\

        \midrule

        Linear
        \\Zeeman
        & $0$
        & $0.001$
        & $2\times10^{-18}$
        \\

        \midrule

        \begin{tabular}{l}
            $2^{\rm nd}$ order\\
            Zeeman
        \end{tabular}
        & $1366.850$
        & $5\times10^{-3}$
        & $1.2\times10^{-20}$
        \\

        \midrule

        \begin{tabular}{l}
            Quadratic\\
            Stark shift\\
            (secular\\
            motion and\\
            IMM)
        \end{tabular}
        & $5.8$
        & $1$
        & $2.4\times10^{-18}$
        \\

        \midrule

        \begin{tabular}{l}
            $2^{\rm nd}$ order\\
            Doppler\\
            (secular\\
            motion and\\
            IMM)
        \end{tabular}
        & $-3.8$
        & $0.7$
        & $1.7\times10^{-18}$
        \\

        \midrule

        \begin{tabular}{l}
            Quadratic\\
            Stark and $2^{\rm nd}$\\
            order Doppler\\
            (excess\\
            micromotion)
        \end{tabular}
        & $0.1$
        & $0.2$
        & $4.8\times10^{-19}$
        \\

        \midrule

        \begin{tabular}{l}
            Black Body\\
            Radiation
        \end{tabular}
        & $355$
        & $10$
        & $2.4\times10^{-17}$
        \\

        \midrule

        \begin{tabular}{l}
            AC Stark\\
            shift (Lasers)
        \end{tabular}
        & $9$
        & $2$
        & $4.8\times10^{-18}$
        \\

        \midrule

        \textbf{Total}
        & $\mathbf{1733}$
        & $\mathbf{10}$
        & $\mathbf{2.5\times10^{-17}}$
        \\

        \bottomrule
    \end{tabular}

    \caption{
        \textbf{Systematic shifts and error for the spectroscopy evaluation
        of the clock.}
        A summary of the most relevant shifts. A frequency offset of
        1.733~Hz with an uncertainty of 10~mHz corresponding to the square
        root of the sum of the squared individual errors is obtained.
    }
    \label{tb:absolutbudget}
\end{table*}

Table~\ref{tb:absolutbudget} summarizes the evaluation of the systematic shifts and error for the absolute frequency via the GNSS link. The gravitational red shift for a clock at rest on the Earth's surface can be calculated with \cite{mehlstaubler2018atomic}:
\begin{equation}
\frac{\Delta \nu_{\textrm{GR}}}{\nu_{0}}=\frac{gH}{c^{2}}
\end{equation}
where $g$ is the gravity acceleration and $H$ is the height of the clock respect to the zero potential. To determine $H$ for our clock measurements we implemented the GNSS/geoid approach \cite{denker2018geodetic} using the European gravimetric geoid (EGG2015) geoid model and the GNSS coordinates obtained with the antenna and the receiver. To perform the calculations we used the online products provided by the International Service for the Geoid \cite{ISG}. To these calculations we added the measurement of the height difference between the antenna (located on the roof) and the ion trap (located in the -1$^{st}$ floor) using a laser distance sensor. We estimate that the height of the clock above the geoid is $H =579 (1)$~m, assuming a conservative error of 1~m. The gravitational red shift is equal to: $(\Delta \nu_{GR}=25.95\pm0.05)$~Hz. The shift and uncertainty of the UTC(PTB) with respect to the TAI and of the TAI with respect to the TT can be found in the report of the BIPM \cite{circularTBIPM, circularTTBIPM}. The transition frequency measurement during the 2021 campaign is 411 042 129 776 400.4~Hz with an uncertainty of 0.6~Hz.

\begin{table*}
    \centering
    \small
    \setlength{\tabcolsep}{8pt}
    \renewcommand{\arraystretch}{1.05}

    \begin{tabular}{cccc}
        \toprule
        \textbf{Effect} &
        \textbf{Shift (Hz)} &
        \textbf{Error (Hz)} &
        \begin{tabular}{c}
            \textbf{Fractional}\\
            \textbf{error}
        \end{tabular}
        \\
        \midrule

        \begin{tabular}{c}
            Systematic\\
            shift
        \end{tabular}
        &
        $1.73$
        &
        $0.01$
        &
        $2.5\times10^{-17}$
        \\

        \begin{tabular}{c}
            Gravitational\\
            shift
        \end{tabular}
        &
        $25.9$
        &
        $0.05$
        &
        $1\times10^{-16}$
        \\

        \begin{tabular}{c}
            UTC(PTB)\\
            referenced to\\
            TAI
        \end{tabular}
        &
        $-0.04$
        &
        $0.12$
        &
        $2.9\times10^{-16}$
        \\

        \begin{tabular}{c}
            TAI\\
            referenced to\\
            TT(SI)
        \end{tabular}
        &
        $-0.08$
        &
        $0.06$
        &
        $1.4\times10^{-16}$
        \\

        \midrule

        \textbf{Total}
        &
        $\mathbf{27.51}$
        &
        $\mathbf{0.14}$
        &
        $\mathbf{3\times10^{-16}}$
        \\

        \begin{tabular}{c}
            Frequency\\
            referenced to\\
            H-maser
        \end{tabular}
        &
        \begin{tabular}{c}
            $411\,042\,129$\\
            $776\,427.9$
        \end{tabular}
        &
        $0.6$
        &
        $1.5\times10^{-15}$
        \\

        \midrule

        \textbf{Absolute frequency}
        &
        \begin{tabular}{c}
            $\mathbf{411\,042\,129}$\\
            $\mathbf{776\,400.4}$
        \end{tabular}
        &
        $\mathbf{0.6}$
        &
        $\mathbf{1.5\times10^{-15}}$
        \\

        \bottomrule
    \end{tabular}

    \caption{
        \textbf{Systematic shifts and error for the absolute frequency via GNSS link.}
        The table gathers all the shifts with their respective errors described
        in the manuscript. The transition frequency measurement during the
        2021 campaign is $411\,042\,129\,776\,400.4$~Hz with an uncertainty
        of $0.6$~Hz.
    }
	\label{tb:errorbudget}
	\end{table*}



\begin{figure}[H]
	\centering
	\includegraphics[width=1\linewidth]{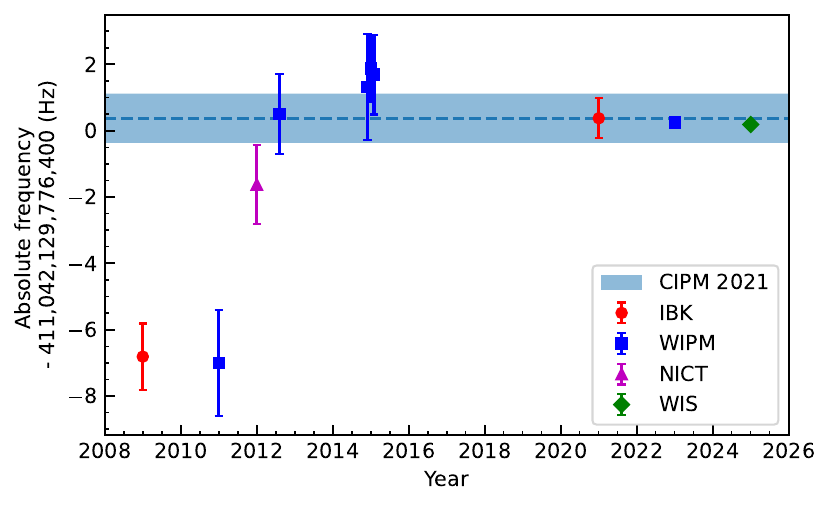}
	\caption{\textbf{Overview of absolute frequency measurements of the $^{40}$Ca$^{+}$ clock transition-} The measurement result from our campaign (IBK '21) is in agreement with the latest measurements of the Wuhan group (WIPM '12, WIPM '14-'15 and WIPM'23) \cite{huang2017comparison,huang2016frequency, zhang2023absolute} and Weizmann group (WIS'25) \cite{sokolik2026direct}. They are however in disagreement with the three first measurements performed between 2009 and 2012 \cite{chwalla2009absolute,matsubara2012direct,huang2012hertz}. As a reference the recommended frequency and uncertainty by the CIPM (International Committee for Weights and Measures) in 2021 is shown in light blue \cite{margolis2024cipm}.}
	\label{fig:Cahistory}
\end{figure}

\section{Conclusions}
In this work we presented the absolute frequency of the $4s$ $ ^{2}S_{1/2}\leftrightarrow 3d$ $^{2}D_{5/2}$  $^{40}$Ca$^{+}$ transition with respect to the UTC(PTB) to be (411 042 129 776 400.4$\pm0.6$)~Hz. The fractional uncertainty of 1.5$\times 10^{-15}$ is mainly limited by the instability of the passive hydrogen maser. The stability of the link could be greatly improved by the implementation of a more stable source such as an active hydrogen maser. Nonetheless, the establishment of the GNSS link in our laboratory represents a major milestone: it offers the possibility to compare our ion clocks to primary and secondary frequency standards in the world. These comparisons are a fundamental tool to facilitate consistency checks of the performance and accuracy of our optical clocks \cite{2020direct}. Frequency comparisons using fiber links have proven to outperform satellite-based links showing residual instabilities of few parts in 10$^{19}$ \cite{droste2013optical, bercy2014line}. However, the implementation of fiber link connections between remotes sites is still impractical in some geographical regions, such as Innsbruck. Satellite links offer a simple solution without requiring too much equipment. The implementation of a passive hydrogen maser, even if not ideal, represents also a more affordable and practical solution as a stable reference source to perform these frequency comparisons. As illustrated in Figure~\ref{fig:Cahistory}, our measurements are in agreement with the latest measurement of the Gao group but disagree with the first measurements performed by our group.
We presented also a new revised measurement of the Land\'e g-factor of the $3d$ $^{2}D_{5/2}$ level to be g$_{5/2}= 1.200329(1)$. The main difference introduced for these measurement with respect to previous ones is the inclusion of the effects of the oscillating magnetic-field arising from ion trap-induced rf currents in electrodes. Our measurements are mainly limited by the uncertainty in the determination of the transversal oscillatory field, which could be reduced by carrying out the experiments at a reduced value of the dc-magnetic field defining the quantization axis.  

We acknowledge funding by the European Space Agency (contract number 4000102396/10/NL/SFe) and by the Institut f\"ur Quanteninformation GmbH. We would also like to thank Andreas Bauch and Thomas Polewka from the Dissemination of time group from the Time and Frequency department at PTB for their help with the frequency comparison measurements and helpful insights. MKJ acknowledges support from the Singapore University of Technology and Design and the MOE Tier 1 grant (Grant No. SAP $2026\_001$) during the final stage of manuscript preparation. The raw data used in the present studies are available on https://zenodo.org/records/23005683. 

\bibliography{ref}

\end{document}